\documentclass[aps,prl,twocolumn,superscriptaddress,showpacs,letterpaper,reprint]{revtex4-1}

\usepackage{graphicx}
\usepackage{dcolumn}
\usepackage{bm}
\usepackage{amsmath}
\usepackage{multirow}

\begin{document}

\newcommand{\mevcc}{\!\mathrm{MeV}\!/c^2}
\newcommand{\mevc}{\!\mathrm{MeV}/\!c}
\newcommand{\mev}{\!\mathrm{MeV}}
\newcommand{\gevcc}{\!\mathrm{GeV}/\!c^2}
\newcommand{\gevc}{\!\mathrm{GeV}/\!c}
\newcommand{\gev}{\!\mathrm{GeV}}

\title{Precision Measurement of the Mass of the $D^{*0}$ Meson and the Binding Energy of the $X(3872)$ Meson as a $D^0\overline{D^{*0}}$ Molecule}

\author{A.~Tomaradze}
\author{S.~Dobbs}
\author{T.~Xiao}
\author{Kamal~K.~Seth}

\affiliation{Northwestern University, Evanston, Illinois 60208, USA}

\date{\today}

\begin{abstract} 
A precision measurement of the mass difference between the $D^0$ and $D^{*0}$ mesons has been made using 316~pb$^{-1}$ of $e^{+}e^{-}$ 
annihilation data taken at $\sqrt{s}=4170$~MeV using the CLEO-c detector. We obtain $\Delta M \equiv M(D^{*0})-M(D^0)
=142.007\pm0.015$(stat)~$\pm$~0.014(syst)~MeV,
as the average for the two decays, $D^0\to K^-\pi^+$ and $D^0\to K^-\pi^+\pi^-\pi^+$. 
The new measurement of $\Delta M$ leads to 
$M(D^{*0})=2006.850\pm0.049$~MeV, and the currently most precise measurement 
of the binding energy of the ``exotic'' meson X(3872) if interpreted as a $D^0D^{*0}$ hadronic molecule, 
$E_{b}(\text{X}(3872))\equiv M(D^0D^{*0})-M(\text{X}(3872))=3\pm192$ keV.
\end{abstract}

\maketitle

Of all the claims and counterclaims for the so-called ``exotic'' mesons,
which do not fit in the pictures of conventional $q\bar{q}$ 
mesons~\cite{isgure},
the most intriguing one is X(3872). Its existence has been confirmed from
numerous measurements, by Belle~\cite{2}, CDF~\cite{3}, D0~\cite{4}, 
BaBar~\cite{5}, LHCb~\cite{6} and CMS~\cite{7}, and its mass, width, and
spin are respectively, $M(\text{X}(3872))=3871.69\pm0.17$~MeV,
$\Gamma(\text{X}(3872))<1.2$~MeV, and $J^{PC}=1^{++}$~\cite{1}.
Although many different suggestions for the structure of X(3872) 
exist in the literature~\cite{9,10,11,12}, the closeness of the 
X(3872) mass to the sum of the masses of the $D^0$ and  $D^{*0}$ mesons,
and the smallness of its width have made the suggestion that it is a weakly 
bound hadronic molecule made of the $D^0$ and  $D^{*0}$ mesons extremely
attractive~\cite{20}. To submit this provocative suggestion to experimental
test it is important to measure the binding energy of X(3872), indeed
to determine if it is bound at all. This Letter reports on the results of 
just such a measurement.
Throughout this Letter we use the PDG~\cite{1} convention for units, with masses in MeV and momenta in MeV/$c$, and inclusion of charge-conjugate states is implied.

A measurement of the binding energy of X(3872) requires the knowledge
of three masses, $M(\text{X}(3872))$, $M(D^0)$, and $M(D^{*0})$, with the most accurately determined value of $M(D^{*0})$
 obtained by measuring the mass difference,
 $\Delta M\equiv M(D^{*0})-M(D^0)$. Since the discovery of X(3872) 
in 2003, the precision in the value of the mass of the X(3872) has steadily improved 
from $\pm800$~keV, originally, to the present average with error of 
$\pm170$~keV~\cite{1} because of numerous 
improved measurements. Similarly, the precision of the value  $M(D^0)$ 
has improved, from $\pm1000$~keV, originally, to $\pm180$~keV by a CLEO
measurement of $M(D^0)$  in 2007~\cite{d0pub}, and to  $\pm40$~keV due to two recent higher-precision  
measurements of $M(D^0)$ by BaBar~\cite{21}, and our recent publication~\cite{22}. 
As a consequence, the determination of the binding energy of $\text{X}(3872)$
 as a $D^0\overline{D^{*0}}$ molecule has changed from 
($600\pm600$) keV in 2007~\cite{d0pub} to ($126\pm204$) keV in 2014~\cite{22}.

Through  all these improvements, the mass difference  $\Delta M$ has remained
fixed at the value,  $\Delta M=142.120\pm0.070$~MeV as measured by CLEO in
1992 using data taken at the $\Upsilon(4S)$ resonance~\cite{23}. 
To determine the binding energy of  X(3872) with the
 highest possible precision, it has become
imperative to make a new higher-precision measurement of  
$\Delta M \equiv M(D^{*0})-M(D^0)$. In this Letter we report on such a 
measurement using data taken at the $\psi(4160)$ resonance, which decays
into $D^{*}\overline{D^*}$, $D^{*}\overline{D}$, and $D\overline{D}$.  
We analyze $D^{*0}\to D^0\pi^{0}$, and  $D^0$ decays, 
$D^0 \rightarrow K^{-} \pi^{+}$ (henceforth $K\pi$), and
 $D^0\to K^{-}\pi^{+}\pi^{+}\pi^{-}$ (henceforth $K3\pi$). 

We use 316~pb$^{-1}$ of $e^+e^-$ annihilation data taken at 
$\sqrt{s}=4170$~MeV with the CLEO-c detector.
The CLEO-c detector \cite{CLEOcDetector} consists of a CsI(Tl)
electromagnetic calorimeter, an inner vertex drift chamber, a central drift
chamber, and a ring-imaging Cherenkov (RICH) detector, all inside a
superconducting solenoid magnet providing a 1.0 Tesla magnetic field. The
acceptance for charged and neutral particles is $|\cos\theta|<0.93$. Charged-particle 
momentum resolution is $\sigma_p/p = 0.6\%$~@~1~GeV/$c$.  Photon energy resolution
is $\sigma_E/E=2.2\%$~@~1~GeV, and $5\%$~@~100~MeV. 
The detector response was studied using a GEANT-based~\cite{GEANTMC} Monte Carlo
simulation.

We select events with
well-measured tracks by requiring that they be fully contained in the
barrel region ($|\cos\theta| < 0.8$) of the detector, and have transverse
momenta $>120$~MeV/$c$.

In our previous article on the precision measurement of the mass of the $D^0$ meson~\cite{22}, we made a
precision recalibration of the CLEO-c 
solenoid  magnetic field and determined a  correction of
$(2.9\pm0.4)\times10^{-4}$ in the default calibration of the CLEO-c magnetic field. The data we use in the present investigation were taken just after this
recalibration.  We use the same corrected field as determined in our previous article  in the present Letter.

\begin{figure}[!t]
\includegraphics*[width=3.5in]{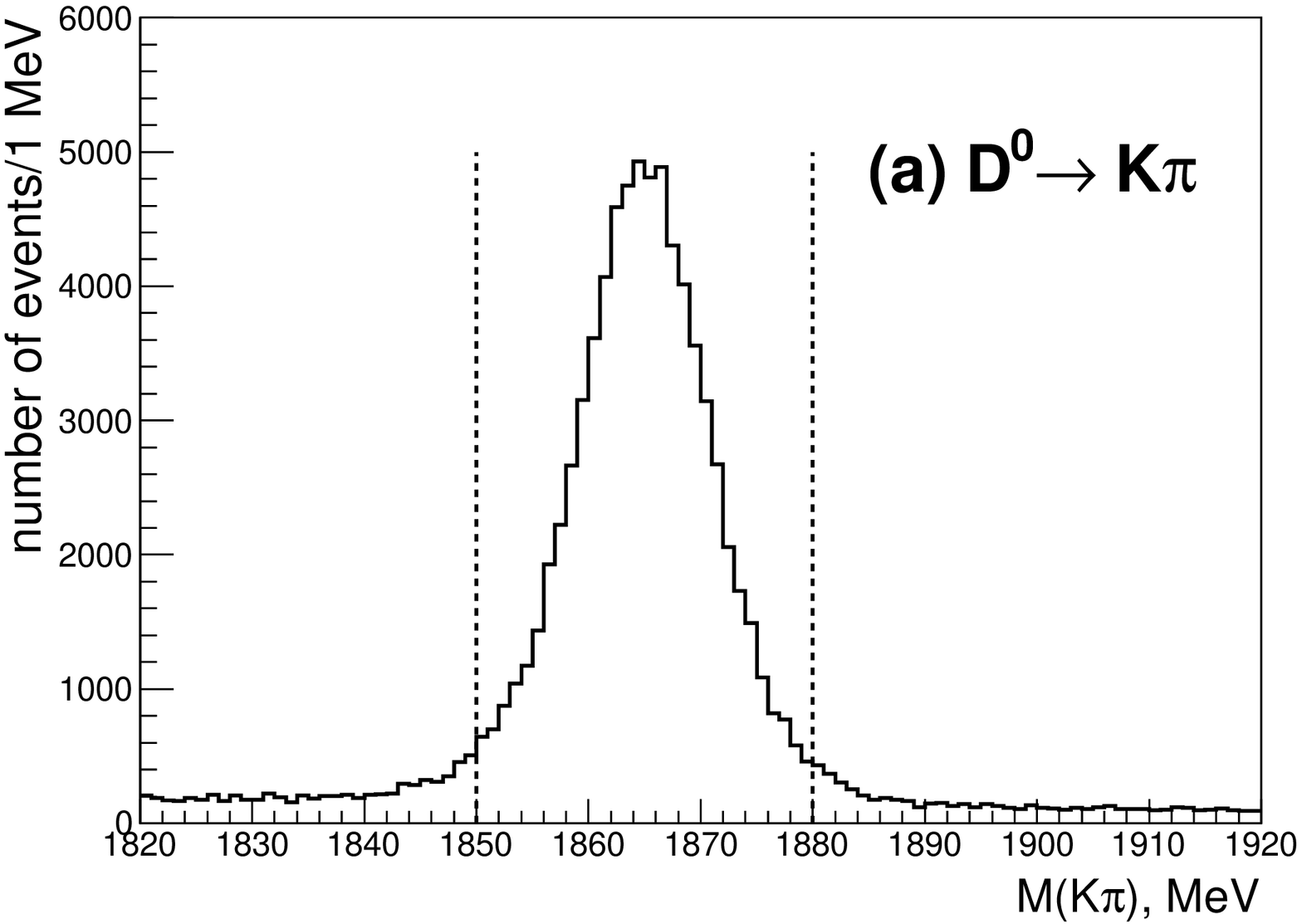}
\includegraphics*[width=3.5in]{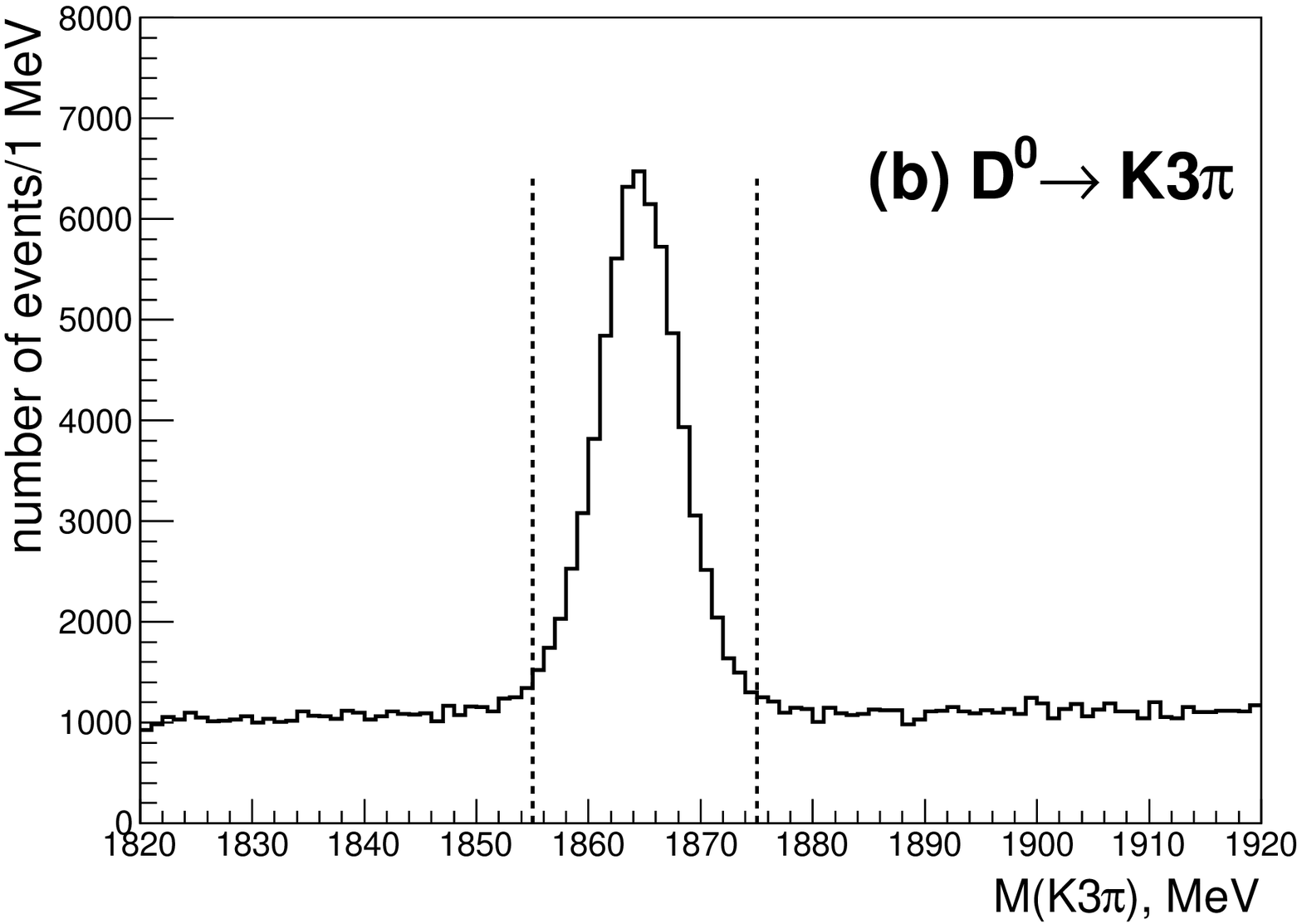}
\caption{Reconstructed mass spectra for the candidates for the decays
(a) $D^0\to K\pi$,
(b) $D^0\to K3\pi$.  
The vertical dashed lines show the regions in which we accept $D^0$ mesons,
 $M(D^0_\text{cand})=1850-1880$~MeV, and
$M(D^0_\text{cand})=1855-1875$~MeV, for  $D^0 \rightarrow K \pi$  and
$D^0\to K3\pi$ channels, respectively.}
\end{figure}

\begin{figure}[!t]
\includegraphics*[width=3.5in]{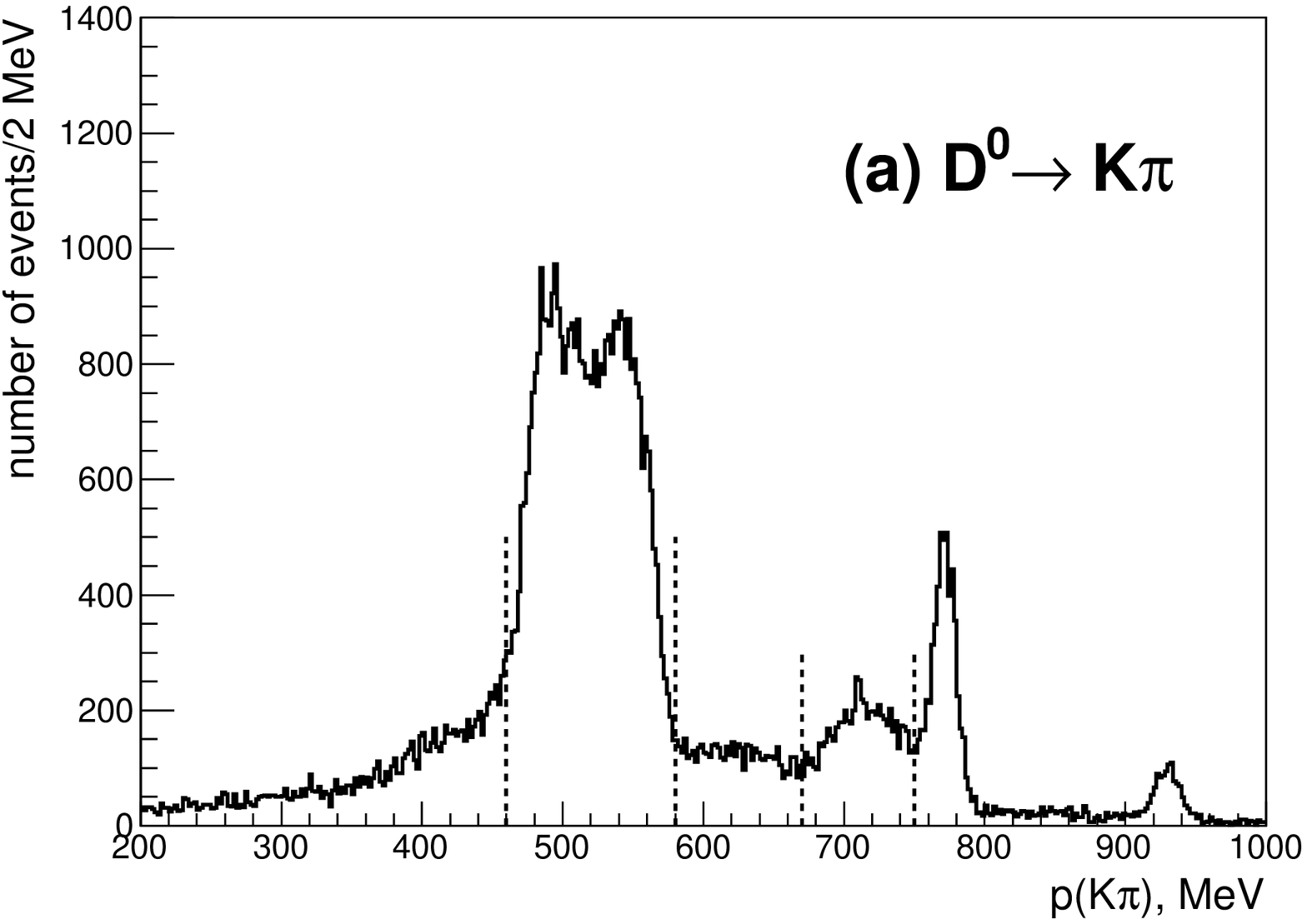}
\includegraphics*[width=3.5in]{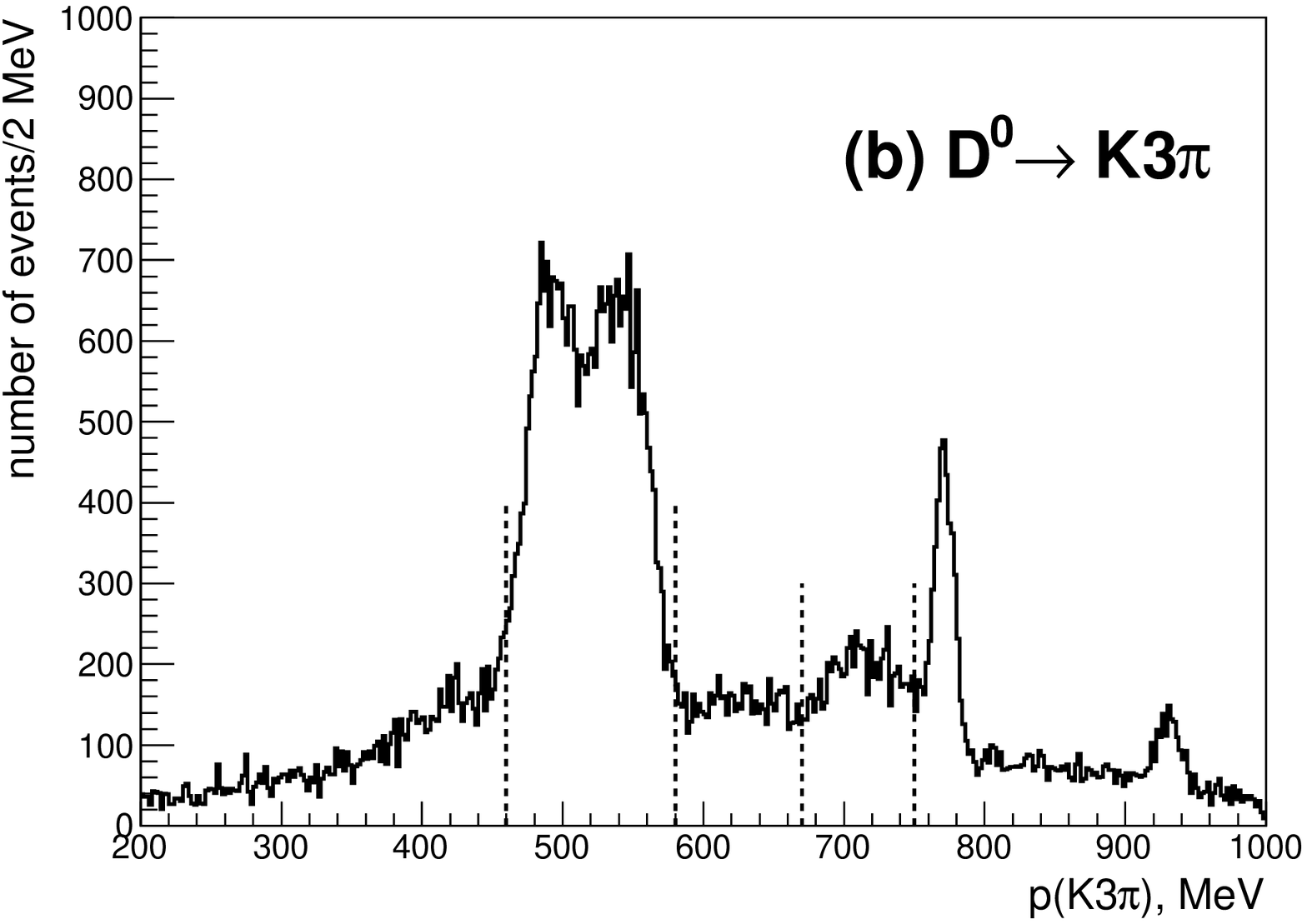}
\caption{Momentum spectra in data for (a)  $D^0 \rightarrow K \pi$
and (b) $D^0\rightarrow K3\pi$ for the candidates identified in Fig.~1.
 As described in the text, the enhancements correspond to momenta for the different expected final states with two charm mesons.
The vertical dashed lines define the intervals of  $p(D^0_\text{cand})$ in which
the events were accepted for analysis.} 
\end{figure}

Charged pions and kaons were identified using information from both 
the drift chamber $dE/dx$ and the RICH detector.  First, it was required that the $dE/dx$ of the charged particle track be consistent within $3\sigma$ of the respective pion or kaon hypothesis.  For tracks with momenta~$>700$~MeV, for which information from the RICH detector is available, the log-likelihood $L_i^{RICH}$, as described in Ref.~\cite{rich}, was constructed.  For such tracks, the combined $dE/dx$ information $\sigma_i^{dE/dx}$,
 and the log-likelihood $L_i^{RICH}$ are used to distinguish between particle hypotheses $i$ and $j$,
\begin{equation}
\Delta L_{i,j} = (\sigma_i^{dE/dx})^2 - (\sigma_j^{dE/dx})^2 + L_i^{RICH} - L_j^{RICH}.
\end{equation}
For tracks with momenta $<700$~MeV, for which RICH information is not 
available, 
\begin{equation}
\Delta L_{i,j} = (\sigma_i^{dE/dx})^2 - (\sigma_j^{dE/dx})^2.
\end{equation}

To identify pions, it was required that $\Delta L_{\pi,K}<0$.  For kaons, it was required that $\Delta L_{\pi,K}>0$. 

We reconstruct $\pi^{0}\to\gamma\gamma$ decays 
using photons only in the barrel region, $|\cos \theta| \le 0.80$.
Photon candidates are defined as calorimeter showers with
$E_\gamma(\mathrm{barrel})>30$ MeV and a transverse energy spread consistent
with that of an electromagnetic shower.
The photon candidates from $\pi^0$ decays
are required to have a two-photon invariant-mass, $M(\gamma\gamma)$, within $\pm15$~MeV
of the nominal $M(\pi^{0})=135.0$~MeV~\cite{1}, and to have neither photon candidate
combine with another photon candidate in the event to obtain an invariant mass
closer to $M(\pi^0)$.
These candidates were kinematically fit with $M(\gamma\gamma)$ constrained
to the nominal $\pi^0$ mass in order to improve the energy resolution.

We reconstruct single $D^{*0}$ candidates through the decays 
$D^{*0} \to \pi^0 D^0$, $D^0 \rightarrow K \pi$ and
 $D^0\to K3\pi$, by first reconstructing the $D^0$ decay and then looking for the 
corresponding $\pi^0$ to make $D^{*0}$.

We select $D^0$ candidates using the standard CLEO D--tagging criteria,
which impose very loose requirements on the beam-energy-constrained
$D^0$ mass, as described in Ref. \cite{dtag}. 

In Fig.~1, we show the reconstructed invariant-mass spectra for $D^0 \rightarrow K\pi$  and 
$D^0\to K3\pi$  candidates.

At $\sqrt{s}=4170$~MeV, $D^0$ mesons can be produced through several different decays: 
$e^+e^- \to D{^*}\overline{D^*}$, ${D^*}\overline{D}$, and $D\overline{D}$. As illustrated in Fig.~2, these different decays have different yields, and they populate different ranges of $D^0$ momenta.  The enhancement in $D^0$ momenta at $\approx500$~MeV arises from $e^+e^-\to D^{*0}\overline{D^{*0}}$, and decays of both $D^{*0}$ and $\overline{D^{*0}}$ to the corresponding $D^0$ and $\overline{D^0}$.  The enhancement at $\approx720$~MeV arises from $e^+e^-\to D^{*0} \overline{D^0}$, and $D^{*0}$ decaying to $D^0$.  The enhancement at $\approx950$~MeV arises from $e^+e^- \to D^0\overline{D^0}$.
GEANT-based~\cite{GEANTMC} generic Monte Carlo is found to faithfully reproduce both the
observed enhancement, and the background, which arises from the other decays 
of $D$ mesons. 
 Events  of interest for the present analysis are those in which at 
least one $D^{*0}$ (or $\overline{D^{*0}}$) is formed.
In the momentum 
distributions of Fig.~2, there is an enhancement centered 
at  $p(D^0_\text{cand})\approx520$ MeV, which arises from events with both $D^{*0}$ and  
$\overline{D^{*0}}$, which decay into  $D^{0}$ and $\overline{D^{0}}$, respectively, 
and an enhancement at  $p(D^0_\text{cand})\approx720$ MeV which arises 
from events with  only one  $D^{*0}$ or $\overline{D^{*0}}$.
The yield in the enhancement at  $p(D^0_\text{cand})\approx520$ MeV is, as expected, nearly a factor four larger 
than that in the  enhancement at  $p(D^0_\text{cand})\approx720$ MeV.

We reconstruct  $D^{*0}$ (or  $\overline{D^{*0}}$) candidates for the events
in the enhancements at  $p(D^0_\text{cand})\approx520$ MeV and  $p(D^0_\text{cand})\approx720$ 
MeV, and construct the distributions for $\Delta M\equiv M(D^{*0}_\text{cand})-M(D^0_\text{cand})$.
These are shown separately for the decays $D^0 \rightarrow K\pi$ and
 $D^0 \rightarrow K3\pi$  in Fig.~3.

\begin{table*}[!tb]
\caption{The results of the fits.}
\begin{center}
\begin{tabular}{lcc}
\hline \hline
Fit      & $D^0\to K\pi$ (Fig. 3(a))   &  $D^0\rightarrow K3\pi$  (Fig. 3(b)) \\
\hline
Number of signal events & $6344\pm178$ & $3697\pm134$ \\ 
FWHM of $\Delta M$ distribution (MeV) & $\approx2.7$ & $\approx2.6$ \\
$\chi^2/d.o.f$ of 250~keV binned fit & 1.10 & 1.02 \\
$\Delta M$ from unbinned fit (MeV) & $142.007\pm0.016(\text{stat})$ & $142.008\pm0.024=7(\text{stat})$ \\
\hline \hline
\end{tabular}
\end{center}
\end{table*}

\begin{figure}[!t]
\includegraphics*[width=3.5in]{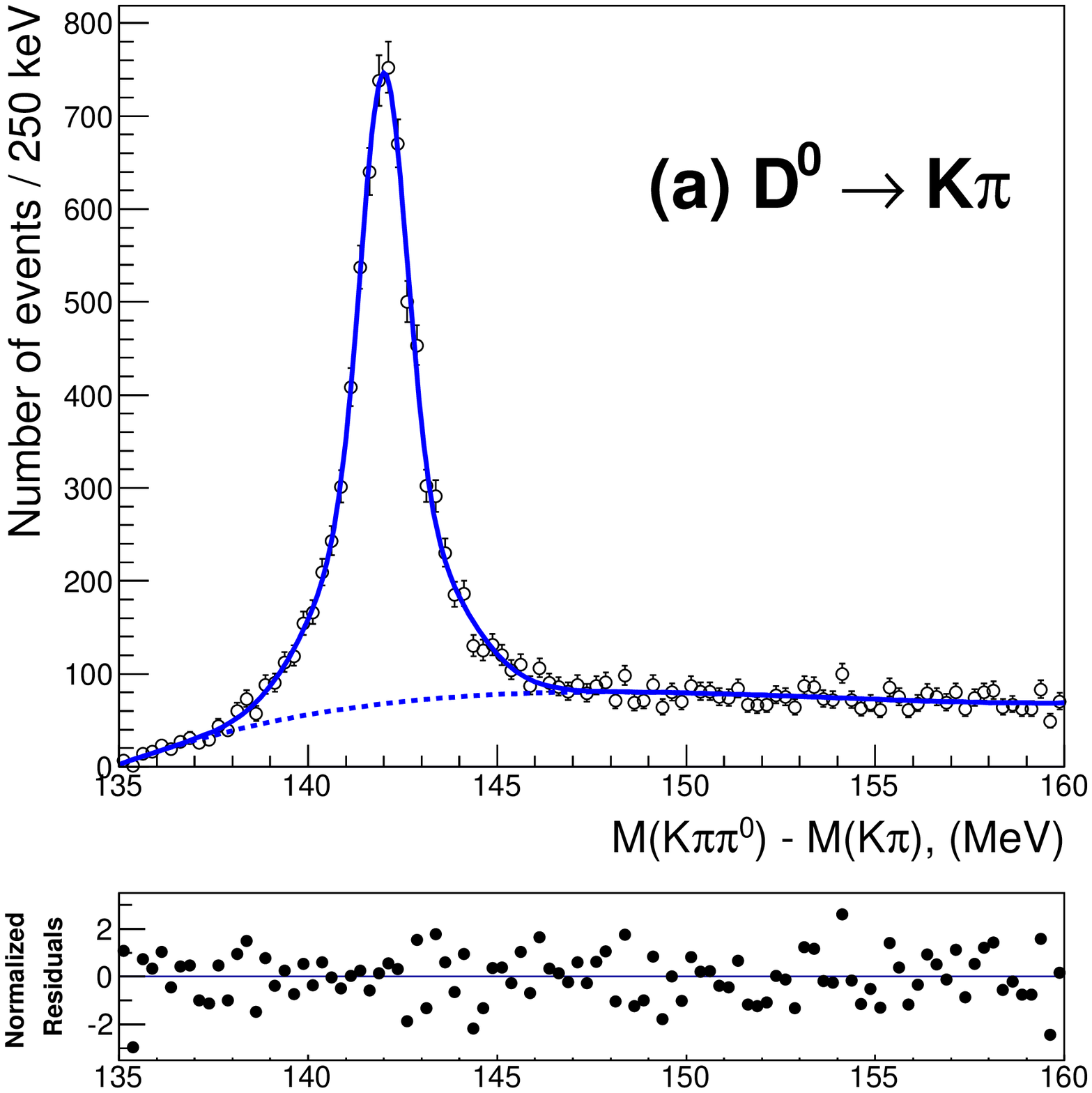}
\includegraphics*[width=3.5in]{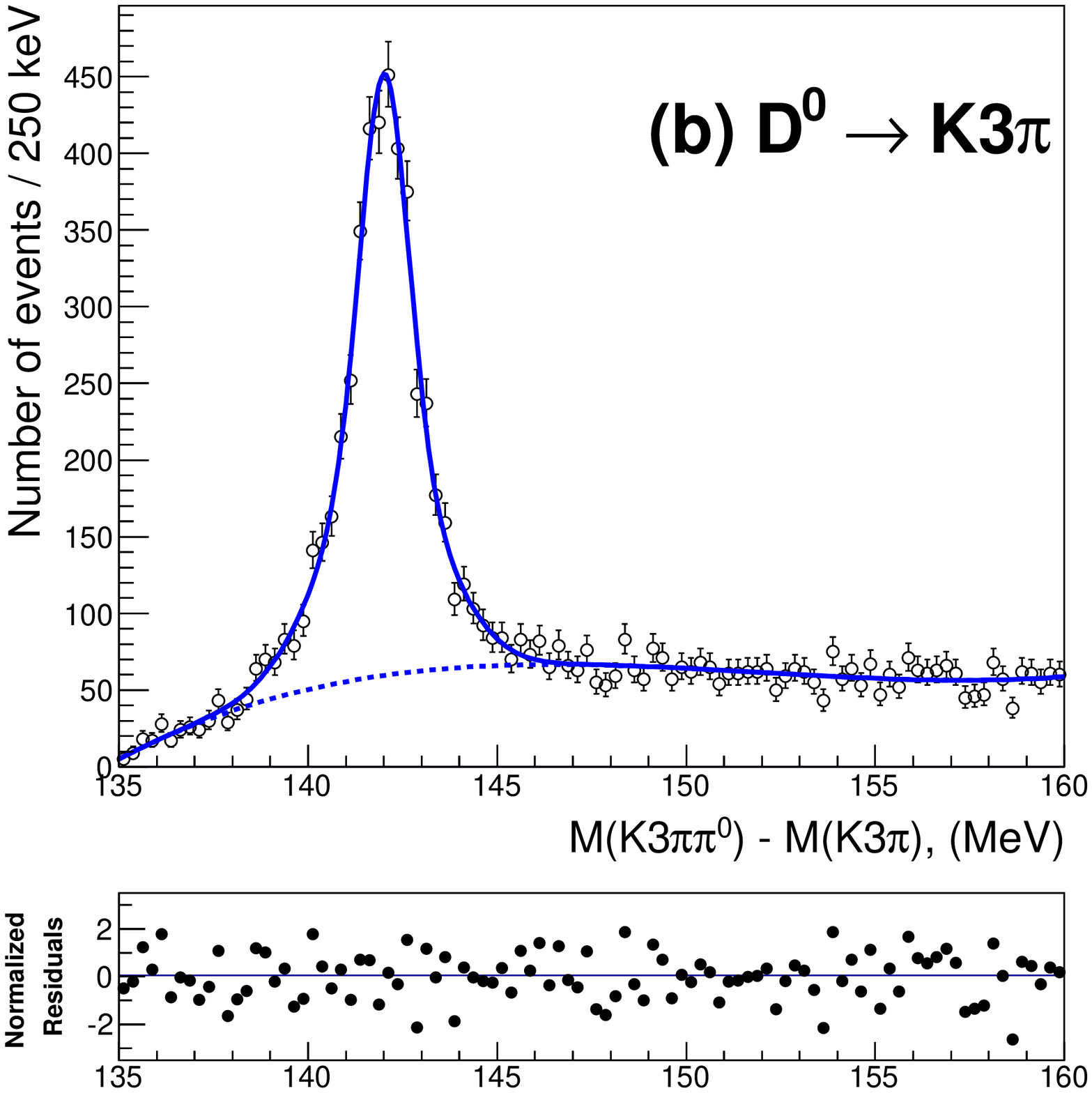}
\caption{The $\Delta M$ spectra and fits for the decays
(a) $D^0\to K\pi$, and
(b)  $D^0\rightarrow K3\pi$. The fits (see text) were done to the unbinned data 
distributions, but for clarity the fits and corresponding residuals are shown for 250~keV bins.}
\end{figure}

Since most of the instrumental uncertainties cancel
in the measurement of the mass difference $\Delta M$, the resolution width of
$\Delta M$ is approximately a factor of six smaller (FWHM $\approx2.7$~MeV) than that for
the individual $D^0$ and $D^{*0}$ masses (FWHM $\approx17$~MeV).

\begin{table*}[!tb]
\caption{Systematic errors in $\Delta M \equiv M(D^{*0}_\text{cand})-M({D^0_\text{cand}})$ in keV.}
\begin{center}
\begin{tabular}{lccc}
\hline \hline
Source      & $D^0\to K\pi$   &  $D^0\rightarrow K3\pi$  \\
\hline
Photon energy calibration $(\pm0.4\%)$             & 10        & 10 \\
Charged particle momentum calibration (B-field $\pm0.4\times10^{-4}$)            & 3        & 3 \\
Signal shape (default vs. single Gaussian)                  & 10        & 7 \\
Background shape (polynomial order +1)              & 1        & 7 \\
$K^{\pm}$ mass ($\pm16$~keV)                & 3        & 3 \\
Cut variations in $M(D^0_\text{cand})$, Fig.~1, ($\pm2.5$~MeV) & 4       & 1 \\
Cut variations in $p(D^0_\text{cand})$, Fig.~2, ($\pm10$~MeV) & 4       & 1 \\
Cut variations in $M(\gamma\gamma)$ ($\pm2$~MeV)  & 4       & 6 \\
 &  &  \\
Total                          & 16       & 16 \\
\hline \hline
\end{tabular}
\end{center}
\end{table*}

For both $K\pi$ and $K3\pi$, we fit the unbinned spectra to
backgrounds parameterized by a second-order polynomial and signal
parametrized as the sum of a Gaussian 
and another Gaussian with the same mean, but different widths on each side of the mean.
Good fits are obtained.

For clarity, we show the $\Delta M$ spectra for $D^0\to K \pi$ (Fig.~3(a)) and $D^0 \to K3\pi$ (Fig.~3(b)), and the corresponding plots of residuals for the data binned in 250~keV bins.  The results of the fits are listed in Table~I.  As is customary, the peak values from the unbinned fits are assumed to be measures of the corresponding mass differences, which are
\begin{eqnarray}
\Delta M(K\pi)  = & 142.007 \pm 0.018\mathrm{~MeV(stat)}, \\
\Delta M(K3\pi) = & 142.008 \pm 0.027\mathrm{~MeV(stat)}.
\end{eqnarray}

We have analyzed the  $\Delta M$ spectra for events in the enhancements at 
$p(D^0_\text{cand})\approx520$ MeV, and  $p(D^0_\text{cand})\approx720$ MeV, separately, and find 
that the results for $\Delta M$ agree with those in Eqs. 3 and 4 within $20-50$~keV, consistent with their statistical uncertainties.

We estimate systematic uncertainties in our results for $\Delta M$ from the following sources.  The results are listed in Table~II.
The CLEO energy calibration for photons is based on the known 
$\pi^{0}$ mass, and on photon energies extracted in the radiative 
decays $\psi(2S)\to \gamma\chi_{cJ}(J=1,2)$, all of which are known with 
high precision. The uncertainty in this calibration is estimated to be to $\pm0.4\%$.
By rescaling the $\pi^{0}$ photon energies by $\pm0.4\%$, we determine the resulting uncertainty in 
$\Delta M$.  In Ref.~\cite{22}, we made a precision 
determination of the CLEO solenoid B-field with an uncertainty of 
factor $0.4\times10^{-4}$.
This results in $\pm3$~keV uncertainty in 
$\Delta M$. The uncertainty in signal shape is
estimated using making alternate fits of the $\Delta M$ spectra by a
single Gaussian in the restricted range of  $\Delta M=141-143$~MeV. 
The systematic error in background shape was obtained by
increasing the order of the polynomial used in the fit by one unit. 
 The PDG2014 mass of the $K^\pm$ has an  uncertainty of 
$\pm16$~keV~\cite{1}.  It leads to $\pm3$~keV uncertainty 
in  $\Delta M$.
Contributions to the systematic  uncertainties derived from variations of
event selection requirements in $M(D^0_\text{cand})$~(Fig.~1),
$p(D^0_\text{cand})$~(Fig.~2), and $M(\gamma\gamma)$ are seperately listed in Table~II. 
Adding all systematic uncertainties in quadrature, the
total systematic uncertainty is estimated to be $\pm16$~keV in $\Delta M(K\pi)$ and $\Delta M(K3\pi)$.

Our final results for $\Delta M$, including systematic errors, are:
\begin{eqnarray}
\Delta M(K\pi)  = & 142.007\!\pm\!0.018(\mathrm{stat})\!\pm\!0.016\mathrm{(syst)~MeV,}~~ \\
\Delta M(K3\pi) = & 142.008\!\pm\!0.027(\mathrm{stat})\!\pm\!0.016\mathrm{(syst)~MeV.}~~
\end{eqnarray}
The results of the two $D^0$ decay final states are in good agreement
(the agreement of the central values within 1~keV is fortuitous).
The weighted average of the 
two determinations, taking proper account of the correlations in the 
systematic errors in the two decay modes, is
\begin{equation}
\Delta M=142.007\pm0.015\mathrm{(stat)}\pm0.014\mathrm{(syst)~MeV}.
\end{equation}

The PDG2014~\cite{1} average of $\Delta M$ based on the 1992 measurement 
of Ref. 17 is $\Delta M=142.12\pm0.07$~MeV based on a total of 1176 counts in 
$K\pi$ and $K3\pi$ decays. Our measurement, based on 10,041 counts, 
has factor $\approx3.5$ smaller overall 
uncertainty, and has 113~keV smaller value of $\Delta M$.

The present measurement (Eq.~7), $\Delta M=142.007\pm0.021$~MeV, and the latest average of $D^0$ mass, 
$M(D^0)=1864.843\pm0.044$~MeV,~\cite{22} lead to
\begin{equation} 
M(D^{*0})=2006.850\pm0.049\mathrm{~MeV}. 
\end{equation}
This compares with the value 
$M(D^{*0})=2006.96\pm0.10$~MeV obtained by the PDG2014~\cite{1} from a simultaneous fit of four masses, $M(D^{*+})$, $M(D^{*0})$, $M(D^{+})$, and $M(D^{0})$.

 Our measured masses lead to
 $M(D^{*0})+M(D^{0})=3871.693\pm0.090$~MeV. Using 
$M(X(3872))=3871.69\pm0.12$~MeV~\cite{1} we obtain the binding energy of $X(3872)$ 
as a proposed $D^0\overline{D^{*0}}$ molecule, 
\begin{eqnarray}
\nonumber E_b\equiv(3871.693\pm0.090)-(3871.69\pm0.17) \mathrm{~MeV}=\\
=3\pm192 \mathrm{~keV.}~~~ 
\end{eqnarray}
The largest contribution to the uncertainty in the above result is due 
to $\pm170$~keV uncertainty in the PDG2014~\cite{1} average value 
of the mass of X(3872).

The negative limiting value of the binding energy $E_b$ implies  that 
$D^0\overline{D^{*0}}$ system could be unbound by as much as 189~keV.
The positive limiting value $E_b=195$~keV  implies that the proposed
 $D^0\overline{D^{*0}}$ molecule, with reduced mass $\mu$, has a \textit{minimum radius}
$R=1/\sqrt{2\mu E_b}$ of 9.9~fm. Hopefully, our new result for the 
binding energy will shed light on the continuing saga of the 
$D^0\overline{D^{*0}}$ molecule and other models of the structure of 
X(3872).

This investigation was done using CLEO data, and as members of
the former CLEO Collaboration we thank it for this privilege.
This research was supported by the U.S. Department of Energy.

\end{document}